\documentclass[preprint,prc,showpacs,showkeys]{revtex4}
\usepackage{amsfonts}
\usepackage{amsmath}
\usepackage{amssymb}
\usepackage{graphicx}
\usepackage{rotating}
\usepackage{amsthm}
\usepackage{txfonts}

\providecommand{\U}[1]{\protect\rule{.1in}{.1in}}
\providecommand{\U}[1]{\protect\rule{.1in}{.1in}}

\begin{document}

\preprint{}
\title{Laser-assisted nuclear photoeffect reexamined\\
}
\author{P\'{e}ter K\'{a}lm\'{a}n$^{1}$\footnote{%
retired from Budapest University of Technology and Economics, Institute of
Physics \newline
e-mail: kalmanpeter3@gmail.com}}
\author{D\'{a}niel Kis$^{2}$}
\author{Tam\'{a}s Keszthelyi$^{1}$}
\affiliation{$^{1}$Budapest University of Technology and Economics, Institute of Physics,
Budafoki \'{u}t 8. F., H-1521 Budapest, Hungary}
\affiliation{$^{2}$Budapest University of Technology and Economics, Institute of Nuclear
Technics, Department of Nuclear Energy, M\H{u}egyetem rkpt. 9., H-1111
Budapest, Hungary}

\begin{abstract}
The S-matrix element and the cross section of the laser-assisted nuclear
photoeffect are recalculated in a gauge invariant manner taking into account
the effect of the Coulomb field of the remainder nucleus. The $\gamma$%
-photon energy dependence of the laser free cross section obtained in the
plane wave and long-wavelength Coulomb-Volkov approximations are compared.
Numerically the laser-assisted partial cross sections with laser photon
energy $2$ $keV$ and some different polarization states of $\gamma $-photon
of energy $3$ $MeV$ are investigated.
\end{abstract}

\volumeyear{year}
\volumenumber{number}
\issuenumber{number}
\eid{identifier}
\date[Date text]{date}
\received[Received text]{date}
\revised[Revised text]{date}
\accepted[Accepted text]{date}
\published[Published text]{date}
\startpage{1}
\endpage{}
\maketitle

\section{Introduction}

The problem of requirement of gauge invariance of perturbation calculus in
matter-field interactions in atomic physics and quantum optics was a central
problem in the 60's and 70's of the last century, and it was finally
satisfactorily clarified in the late 80's \cite{Gauge}, \cite{Lamb}. At this
time the laser assisted x-ray photoeffect was widely investigated and the
gauge invariant calculation of the cross section was also made \cite{KP1}.
In a recent paper \cite{Dadi} the effect of intense coherent electromagnetic
field on the nuclear photoeffect was discussed. The laser-assisted nuclear
photoeffect is a process, which is similar to the laser-assisted x-ray photo
effect (x-ray absorption). Both processes are bound-free transitions of
charged particles (protons and electrons, respectively) that are assisted by
an intense electromagnetic field. In both cases the initial state is
strongly bound so that its change due to the intense laser field may be
neglected. Since the correspondence between the two processes is
straightforward, one can use the results of \cite{KP1} in calculating the
gauge invariant cross section of laser assisted nuclear photoeffect making
the necessary substitutions and modifications in the formulae. 

Therefore we re-discuss the problem of the laser-assisted nuclear
photoeffect satisfying the requirement of gauge invariance and taking into
account the effect of the Coulomb field of the remaining nucleus. The recoil
of \ the remainder nucleus and the initial momentum of the $\gamma $
particle are neglected. The calculation is made in radiation $(pA)$ gauge in
long wavelength approximation (LWA) of the electromagnetic fields. In LWA
the laser field has a vector potential $\overrightarrow{A}_{L}(t)=A_{0}[\cos
(\omega _{0}t)\overrightarrow{e}_{1}-\sin (\omega _{0}t)\overrightarrow{e}%
_{2}]$, that corresponds to a circularly polarized monochromatic wave.%
For the unit vectors $\overrightarrow{e}_{1}$ and $%
\overrightarrow{e}_{2}$ the $\overrightarrow{e}_{1}\cdot \overrightarrow{e}%
_{2}=0$ holds. The corresponding amplitude of the electric field of the
laser is $F_{0}=\omega _{0}A_{0}/c$. The unit vectors of the frame of
reference used are $\overrightarrow{e}_{x}=\overrightarrow{e}_{1}$, $%
\overrightarrow{e}_{y}=\overrightarrow{e}_{2}$ and $\overrightarrow{e}_{z}=%
\overrightarrow{e}_{1}\times \overrightarrow{e}_{2}$, with $\overrightarrow{e%
}_{z}$ in the direction of propagation of the intense field. The vector
potential describing the gamma radiation is $\overrightarrow{A}_{\gamma }=%
\sqrt{2\pi \hbar /\left( V\omega _{\gamma }\right) }\overrightarrow{%
\varepsilon }\exp \left( -i\omega _{\gamma }t\right) $ in the LWA. Here $%
\hbar \omega _{\gamma }$ is the energy and $\overrightarrow{\varepsilon }$
is the unit vector of state of polarization of the $\gamma $ photon, and $V$
is the volume of normalization. The polar angles of $\overrightarrow{%
\varepsilon }$ are $\Theta $ and $\Phi $. The space dependent part of the
initial nuclear (protonic) state $\phi _{0}(\overrightarrow{r})=\left( 2\pi
\right) ^{-1}\beta ^{3/2}e^{-\beta r}/\left(\beta r\right)$, with $\beta
=\nu \sqrt{2mE_{b}}/\hbar $, where $m$ is the rest mass of the proton and $%
-E_{b}$ is its initial energy. The actual values of $E_{b}$ and $\nu $ are $%
E_{b}=0.137$ $MeV$ and $\nu =1.84$ \cite{Dadi}.

\section{Gauge invariant S-matrix element in the LWA Coulomb-Volkov model}

The wave function of a free proton in a repulsive Coulomb field of charge
number $Z$ has the form $\varphi (\overrightarrow{r})=e^{i\overrightarrow{Q}%
\cdot \overrightarrow{r}}\chi (\overrightarrow{Q},\overrightarrow{r})/\sqrt{V%
}$ \cite{Alder}. Here $\overrightarrow{Q}$ is the wave number vector of the
proton. Applying the LWA in $\chi (\overrightarrow{Q},\overrightarrow{r})$,
i.e. taking $\left\vert \chi (\overrightarrow{Q},0)\right\vert =\chi _{C}(Q)$
, 
\begin{equation}
\varphi (\overrightarrow{r})=e^{i\overrightarrow{Q}\cdot \overrightarrow{r}%
}\chi _{C}(Q)/\sqrt{V}  \label{Cb2}
\end{equation}%
with%
\begin{equation}
\chi _{C}(Q)=\left( \frac{2\pi Z\alpha_{f}}{\lambdabar _{p}Q}\right)^{1/2}%
\left[ \exp \left( \frac{2\pi Z\alpha_{f}}{\lambdabar _{p}Q}\right) -1\right]
^{-1/2}.  \label{fjk}
\end{equation}%
Here $\alpha _{f}$ is the fine structure constant and $\lambdabar _{p}=\frac{%
\hbar }{mc}$ is the reduced Compton wavelength of the proton. The solution $%
\left( \ref{Cb2}\right) $ is a LWA of the Coulomb-solution. This
approximation leads to the Fermi correction factor of the decay rate of the $%
\beta $ decay \cite{Blatt}.

For the time dependent state of the proton in the intense field an
approximate non-relativistic solution $\psi $ of the time dependent Schr\"{o}%
dinger equation of a particle in the laser plus Coulomb field is used, which
is called Coulomb-Volkov solution \cite{Coulomb}, \cite{Rosenberg}. For $%
\psi _{\overrightarrow{Q}}\left( \overrightarrow{r},t\right) $ we use the
LWA of nonrelativistic Coulomb-Volkov solution 
\begin{equation}
\psi _{\overrightarrow{Q}}\left( \overrightarrow{r},t\right) =V^{-1/2}e^{i%
\overrightarrow{Q}\cdot \overrightarrow{r}}\chi _{C}(Q)\exp \left( -i%
\widehat{E}t/\hbar \right) f(t)  \label{pszif}
\end{equation}%
with $\widehat{E}=\hbar ^{2}Q^{2}/(2m)+U_{p}$, that is the energy of the
outgoing proton in the intense field, where $U_{p}=e^{2}F_{0}^{2}/(2m\omega
_{0}^{2})$ is the ponderomotive energy. The polar angles of the wave number
vector $\overrightarrow{Q}$ of the outgoing proton are $\vartheta $ and $%
\eta _{0}$. The function $f(t)=\exp [i\alpha \sin (\omega _{0}t+\eta _{0})]$
where $\alpha =\alpha _{\vartheta }\sin \left( \vartheta \right) $ \ with \ $%
\alpha _{\vartheta }=eF_{0}Q/\left( m\omega _{0}^{2}\right) $.

The gauge independent S-matrix element can be obtained with the aid of
Eq.(27) of \cite{KP1} as 
\begin{equation}
S_{fi}=-\frac{\chi _{C}(Q)}{\sqrt{V}}\int \exp [i\left( \widehat{E}%
+E_{b}\right) t/\hbar ]f^{\ast }(t)\frac{\partial }{\partial _{t}}G\left[ 
\overrightarrow{q}\left( t\right) \right] dt,  \label{Sfi}
\end{equation}%
where $G\left( \overrightarrow{q}\right) =\int \phi _{0}(\overrightarrow{r}%
)e^{-i\overrightarrow{q}\cdot \overrightarrow{r}}d^{3}r$ is the Fourier
transform of the initial stationary nuclear state, it is $G\left( 
\overrightarrow{q}\right) =2\sqrt{2\pi \beta }\left( q^{2}+\beta ^{2}\right)
^{-1}$ in our case, and $\overrightarrow{q}\left( t\right) =$ $%
\overrightarrow{Q}-\frac{e}{\hbar c}\overrightarrow{A}$ with $%
\overrightarrow{A}=\overrightarrow{A}_{L}\left( t\right) +\overrightarrow{A}%
_{\gamma }$. Using the $\partial _{t}G=\left( \partial _{q}G\right)
\sum_{j=1}^{j=3}\left( \partial _{A_{j}}q\right) \left( \partial
_{t}A_{j}\right) $ identity $\partial _{t}G=\left( \partial _{q}G\right) 
\frac{e}{\hbar q}\left( \overrightarrow{Q}\cdot \overrightarrow{E}-\frac{e}{%
\hbar c}\overrightarrow{A}\cdot \overrightarrow{E}\right) $ with $%
\overrightarrow{E}=-\frac{1}{c}\partial _{t}\overrightarrow{A}$, i.e. $%
\overrightarrow{E}=\overrightarrow{E}_{L}\left( t\right) +\overrightarrow{E}%
_{\gamma }$.

Now we deal with last factor of $\partial _{t}G$. The $\overrightarrow{Q}%
\cdot \overrightarrow{E}_{L}$ term can be neglected if the pure intense
field induced proton stripping process is negligible since this term
describes the process without the gamma photon. Furthermore, the ratio of
the amplitudes of $\overrightarrow{A}_{\gamma }\cdot \overrightarrow{E}_{L}$
and $\overrightarrow{A}_{L}\cdot \overrightarrow{E}_{\gamma }$ equals $%
\omega _{L}/\omega _{\gamma }\ll 1$. Therefore the $\overrightarrow{Q}\cdot 
\overrightarrow{E}-\frac{e}{\hbar c}\overrightarrow{A}\cdot \overrightarrow{E%
}=\overrightarrow{Q}\cdot \overrightarrow{E}_{\gamma }-\frac{e}{\hbar c}~%
\overrightarrow{A}_{L}\cdot \overrightarrow{E}_{\gamma }$ approximation is
justified to use, where $\overrightarrow{E}_{\gamma }=i\sqrt{2\pi \hbar
\omega _{\gamma }/V}\overrightarrow{\varepsilon }\exp \left( -i\omega
_{\gamma }t\right) $. The relative strength of the $\overrightarrow{Q}\cdot 
\overrightarrow{E}_{\gamma }$ and $\frac{e}{\hbar c}~\overrightarrow{A}%
_{L}\cdot \overrightarrow{E}_{\gamma }$ terms is characterized by the
parameter $\delta =eA_{0}/\left( \hbar cQ\right) $. In the laser free case $%
Q=\sqrt{2m\left[ \hbar \omega _{\gamma }-E_{b}\right] }/\hbar $. Numerical
estimation shows that $\delta \lesssim 0.004$ if $\hbar \omega _{\gamma
}\geq 3$ $MeV$ and $\delta $ increases up to $\delta \simeq 0.05$ if $\hbar
\omega _{\gamma }=0.189$ $MeV$ in the case of laser photon energy and
intensity values discussed in \cite{Dadi}. Therefore the $\overrightarrow{Q}%
\cdot \overrightarrow{E}_{\gamma }$ term is the leading one in the last
factor of $\partial _{t}G$ that after the substitution of the concrete form
of $\partial _{q}G$ results 
\begin{equation}
\frac{\partial }{\partial _{t}}G=\left( \frac{\partial }{\partial q}G\right) 
\frac{e}{\hbar }\frac{\overrightarrow{Q}\cdot \overrightarrow{E}_{\gamma }}{q%
}=-\frac{4\sqrt{2\pi \beta }}{\left[ q^{2}+\beta ^{2}\right] ^{2}}\frac{e}{%
\hbar }\overrightarrow{Q}\cdot \overrightarrow{E}_{\gamma }.  \label{dGt2}
\end{equation}

As to the denominator of $\left( \ref{dGt2}\right) $, the effect of $%
\overrightarrow{A}_{\gamma }$ is negligible in $\overrightarrow{q}(t)$ and
thus $\overrightarrow{q}\left( t\right) =\overrightarrow{Q}-\frac{e}{\hbar c}%
\overrightarrow{A}_{L}$. It was shown above that the amplitude of
oscillation of $\overrightarrow{q}\left( t\right) $ due to the intense field
can be neglected. Moreover the amplitude of oscillation of $\overrightarrow{q%
}\left( t\right) $ compared to $\beta $ is less than $1.4\%$. Therefore $q=Q$
can be used in the denominator of $\left( \ref{dGt2}\right) $.

Using the Jacobi-Anger formula in the Fourier series expansion of $f^{\ast
}(t)$ \cite{Gradstein1} the S-matrix element can be written as 
\begin{equation}
S_{fi}=\sum_{n=n_{0}}^{\infty}\frac{2\pi i}{V}\delta\left[\omega_{n}(Q)%
\right]\chi_{C}(Q)\left(\frac{\partial G}{\partial q}\right)_{q=Q}\frac{e%
\sqrt{2\pi\hbar\omega_{\gamma}}}{\hbar}\mathit{M}_{n}\left(\xi,\alpha\right)
\label{Sfi2}
\end{equation}
with $\mathit{M}_{n}\left(\xi,\alpha\right) =\xi J_{n}(\alpha)e^{-in\eta_{0}}
$, where $\xi =\overrightarrow{Q}\cdot \overrightarrow{\varepsilon }/Q$, $%
J_{n}(\alpha )$ is a Bessel function of the first kind, and 
\begin{equation}
\omega _{n}(Q)=\frac{\hbar Q^{2}}{2m}+\frac{U_{p}+E_{b}}{\hbar }-\omega
_{\gamma }-n\omega _{0}.  \label{OmegaQ}
\end{equation}

\section{Gauge invariant cross section of laser-assisted nuclear photoeffect}

The cross section has the form 
\begin{equation}
\sigma =\sum_{n=n_{0}}^{\infty }\sigma _{n},  \label{Sigma}
\end{equation}
where the partial cross section 
\begin{equation}
\sigma _{n}=\sigma _{n0}\left( Q_{n}\right) \left\vert \mu _{n}\right\vert
^{2}  \label{Sigman}
\end{equation}
with%
\begin{equation}
\sigma _{n0}\left( Q_{n}\right) =\chi _{C}^{2}\left( Q_{n}\right) \alpha _{f}%
\frac{Q_{n}k_{\gamma }}{2\pi \lambdabar _{p}}\left[ \frac{\partial }{%
\partial _{q}}G\left( \overrightarrow{q}\right) \right] _{q=Q_{n}}^{2}.
\label{Sigman0}
\end{equation}%
In our case 
\begin{equation}
\sigma _{n0}\left( Q_{n}\right) =16\alpha _{f}\frac{k_{\gamma }\beta }{%
\lambdabar _{p}}\chi _{C}^{2}\left( Q_{n}\right) \frac{Q_{n}^{3}}{\left[
Q_{n}^{2}+\beta ^{2}\right] ^{4}}.  \label{Sigman20}
\end{equation}%
Here $Q_{n}=\frac{1}{\hbar }\sqrt{2m\left[ \hbar \left( \omega _{\gamma
}+n\omega _{0}\right) -U_{p}-E_{b}\right] }$, $k_{\gamma }=\omega _{\gamma
}/c$. The cases $n<0$ and $n>0$ correspond to laser photon emission and
absorption, respectively. $n_{0}\left( <0\right) $ is the smallest possible
value of $n$, it just fulfills the $\hbar \left( \omega _{\gamma }+n\omega
_{0}\right) -U_{p}-E_{b}>0$ condition ($\left\vert n_{0}\right\vert \simeq
\left( \hbar \omega _{\gamma }-U_{p}-E_{b}\right) /\hbar \omega _{0}$). 
\begin{equation}
\left\vert \mu _{n}\right\vert ^{2}=\int_{0}^{2\pi }\int_{0}^{\pi }\xi
^{2}J_{n}^{2}(\alpha _{n})\sin \vartheta d\vartheta d\eta _{0},  \label{mN20}
\end{equation}%
where $\alpha _{n}=\alpha _{n\vartheta }\sin \vartheta $ \ with \ $\alpha
_{n\vartheta }=eF_{0}Q_{n}/\left( m\omega _{0}^{2}\right) $.

The partial cross section is proportional to the Coulomb factor $\chi
_{C}^{2}(Q_{n})$. The Coulomb factor and therefore also the partial cross
section rapidly decrease with the decrease of $Q_{n}$, i.e. near the
threshold $\left( \hbar \omega _{\gamma }\rightarrow E_{b}\right) $, and it
is the fact in the laser free case too. Therefore near the threshold the
cross section is unobservable.

If the direction of $\overrightarrow{\varepsilon }$ in the gamma flux is
random then $\left\vert \mu _{n}\right\vert ^{2}$ must be averaged as $%
\left\langle \left\vert \mu _{n}\right\vert ^{2}\right\rangle =\int
\left\vert \mu _{n}\right\vert ^{2}\sin \Theta d\Theta d\Phi /\left( 4\pi
\right) $. Applying the spherical harmonics addition theorem, the
orthonormal properties and the sum rule of the spherical harmonics \cite%
{Bethe} one can obtain $\left\langle \xi ^{2}\right\rangle =1/3$ and 
\begin{equation}
\left\langle \left\vert \mu _{n}\right\vert ^{2}\right\rangle
=\int_{0}^{2\pi }\int_{0}^{\pi }\frac{1}{3}J_{n}^{2}(\alpha _{n\vartheta
}\sin \vartheta )\sin \vartheta d\vartheta d\eta _{0}.  \label{mun2}
\end{equation}%
Than the average of the partial cross section $\sigma _{n}$ reads $%
\left\langle \sigma _{n}\right\rangle =\sigma _{n0}\left( Q_{n}\right)
\left\langle \left\vert \mu _{n}\right\vert ^{2}\right\rangle $.

It can be seen from $\left( \ref{mun2}\right) $ that the averaged
differential partial cross section can be written as 
\begin{equation}
\frac{d\left\langle \sigma _{n}\right\rangle }{d\Omega _{Q}}=\sigma _{n0}%
\frac{1}{3}J_{n}^{2}(\alpha _{n\vartheta }\sin \vartheta ),
\label{Sigmandiff}
\end{equation}%
where $d\Omega _{Q}=\sin \vartheta d\vartheta d\eta _{0}$.

In the $n\neq 0$ channels the proton emission vanishes in the direction
parallel $\left( \vartheta =0\text{ or }\vartheta =\pi \right) $ with the
laser beam since $\lim_{x\rightarrow 0}J_{n}\left( x\right) =0$. In the $n=0$
case in this direction $\frac{d\left\langle \sigma _{n}\right\rangle }{%
d\Omega _{p}}=\sigma _{00}\frac{1}{3}$ since $\lim_{x\rightarrow
0}J_{0}\left( x\right) =1$.

In calculating $\left\langle \sigma _{n}\right\rangle $ one can use the $%
J_{-n}(x)=J_{n}(-x)=\left( -1\right) ^{n}J_{n}(x)$ relations to change the $%
\int_{0}^{\pi }d\vartheta $ to $2\int_{0}^{\pi /2}d\vartheta $, and one can
apply the 
\begin{equation}
\int_{0}^{\pi /2}J_{n}^{2}[\alpha _{n\vartheta }\sin \left( \vartheta
\right) ]\sin \left( \vartheta \right) d\vartheta =\frac{1}{\alpha
_{n\vartheta }}\sum_{k=0}^{\infty }J_{2n+2k+1}(2\alpha _{n\vartheta })
\label{Ident1}
\end{equation}%
\cite{Bailey} and the 
\begin{equation}
2\sum_{k=0}^{\infty }J_{2n+2k+1}(2\alpha _{n\vartheta })=\int_{0}^{2\alpha
_{n\vartheta }}J_{2n}\left( x^{\prime }\right) dx^{\prime }  \label{Ident2}
\end{equation}%
\cite{Gradstein1} identities. Using the $x^{\prime }=2\alpha _{n\vartheta }x$
and $dx^{\prime }=2\alpha _{n\vartheta }dx$ change of variable 
\begin{equation}
\left\langle \sigma _{n}\right\rangle =\frac{4\pi }{3}\sigma
_{n0}\int_{0}^{1}J_{2n}\left( 2\alpha _{n\vartheta }x\right) dx.
\label{Signab4}
\end{equation}

In the weak field limit $\left( \alpha _{n\vartheta }\rightarrow 0\right) $
the $\int_{0}^{1}J_{2n}\left( 2\alpha _{n\vartheta }x\right) dx\rightarrow
\int_{0}^{1}J_{2n}\left( 0\right) dx=0$ because $\lim_{x\rightarrow
0}J_{2n}\left( x\right) =0$ in the case of $n\neq 0$ and the $%
\int_{0}^{1}J_{2n}\left( 2\alpha _{n\vartheta }x\right) dx\rightarrow
\int_{0}^{1}J_{0}\left( 0\right) dx=1$ because $\lim_{x\rightarrow
0}J_{0}\left( x\right) =1$ in the $n=0$ case. Therefore the averaged total
cross section $\left\langle \sigma \right\rangle \rightarrow \frac{4\pi }{3}%
\sigma _{00}$. On the other hand $\left\langle \sigma \right\rangle
\rightarrow \sigma _{T}$, that is the total cross section of the laser free
case and with random polarization in the $\gamma $-flux therefore $\sigma
_{00}=\frac{3}{4\pi }\sigma _{T}$.

If we are far from the threshold, i.e. if $\hbar \omega _{\gamma }\gg E_{b}$%
, then the $Q_{n}=Q_{0}$ approximation holds for the intensity and photon
energy parameter pairs ($I=10^{18}$ $W/cm^{2}$ with $\hbar \omega _{0}=2$ $%
eV $; $I=6.25\times 10^{21}$ $W/cm^{2}$ with $\hbar \omega _{0}=100$ and $%
200 $ $eV$; $I=4.0\times 10^{21},1.\times 10^{23}$ and $2.5\times 10^{24}$ $%
W/cm^{2} $ with $\hbar \omega _{0}=2$ $keV$) discussed in \cite{Dadi} and
one can use the $\sigma _{n0}=\sigma _{00}=\frac{3}{4\pi }\sigma _{T}$
substitution in all the above formulae of the partial cross section. Using
this approximation in $\left( \ref{Sigmandiff}\right) $ and applying $%
\sum_{n=-\infty }^{n=\infty }J_{n}^{2}(x)=1$, since $n_{0}$ is a negative
integer of large magnitude, $\frac{d\left\langle \sigma \right\rangle }{%
d\Omega _{Q}}\simeq \frac{1}{4\pi }\sigma _{T}$ and $\left\langle \sigma
\right\rangle \simeq \sigma _{T}$ at the intensities and laser photon
energies discussed.

Now it is supposed that the $\gamma $-photon is polarized. If the
polarization vector $\overrightarrow{\varepsilon }=\overrightarrow{e}_{2}$,
i.e. the polarization vector $\overrightarrow{\varepsilon }$ lies in the
plane of polarization of the circularly polarized laser beam (case $pol,1$)
then $\xi =\overrightarrow{Q}\cdot \overrightarrow{e}_{2}/Q=\sin \vartheta
\sin \eta _{0}$. If the polarization vector $\overrightarrow{\varepsilon }$
of the $\gamma $-photon is $\overrightarrow{\varepsilon }=\overrightarrow{e}%
_{1}\times \overrightarrow{e}_{2}$ (case $pol,2$) then $\xi =\cos \vartheta $%
. Using $\left( \ref{mN20}\right) $ and carrying out the integration over $%
\eta _{0}$ 
\begin{equation}
\left\vert \mu _{n}\right\vert _{pol,1}^{2}=\pi \int_{0}^{\pi
}J_{n}^{2}(\alpha _{0\vartheta }\sin \vartheta )\sin ^{3}\vartheta
d\vartheta ,  \label{mun2pol122}
\end{equation}%
\begin{equation}
\left\vert \mu _{n}\right\vert _{pol,2}^{2}=\int_{0}^{2\pi }\int_{0}^{\pi
}J_{n}^{2}(\alpha _{0\vartheta }\sin \vartheta )\cos ^{2}\vartheta \sin
\vartheta d\vartheta d\eta _{0},  \label{mun2pol2}
\end{equation}%
and $\sigma _{n}$ has the form 
\begin{equation}
\sigma _{n,pol,j}=\frac{3}{4\pi }\sigma _{T}\left\vert \mu _{n}\right\vert
_{pol,j}^{2}.  \label{Sigmanpolj}
\end{equation}%
The total cross sections $\sigma _{pol,j}=\sum_{n_{0}}^{\infty }\sigma
_{n,pol,j}$ can be obtained applying $\sum_{n_{0}}^{\infty
}J_{n}^{2}(x)\simeq 1$ again resulting $\sigma _{pol,1}=\sigma
_{pol,2}=\sigma _{T}$ valid in the cases of the laser intensities and photon
energies discussed.

\section{Numerical results}

\begin{figure}[tbp]
\begin{center}
\resizebox{7.5cm}{!}{\includegraphics*{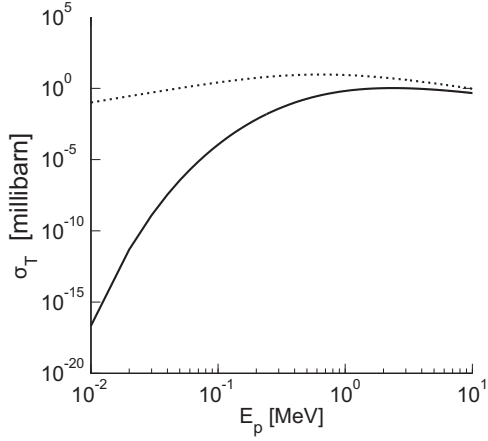}}
\end{center}
\caption{The full line shows the $E_{p}=\hbar \protect\omega _{\protect%
\gamma }-E_{b}$ dependence of the laser free averaged total cross section
given by $\protect\sigma _{T}=\frac{4\protect\pi }{3}\protect\sigma _{00}$
(for $\protect\sigma _{00}$ see $\left( \protect\ref{Sigman20}\right) $ with 
$Q_{0}$). $E_{p}$ is the kinetic energy of the outgoing proton, $\hbar 
\protect\omega _{\protect\gamma }$ is the energy of the $\protect\gamma $
photon and $E_{b}=0.137$ $MeV$ is the binding energy of the proton initially
bound in $^{8}B$. For comparison the $\frac{4\protect\pi }{3}\protect\sigma %
_{00}/\protect\chi _{C}^{2}\left( Q_{0}\right) $ is also plotted as a dotted
line.}
\label{figure1}
\end{figure}

First the $E_{p}=\hbar \omega _{\gamma }-E_{b}$ dependence of the laser free
averaged total cross section is investigated. $E_{p}$ is the kinetic energy
of the outgoing proton, $\hbar \omega _{\gamma }$ is the energy of the $%
\gamma $ photon and $E_{b}=0.137$ $MeV$ is the binding energy of the proton
initially bound in $^{8}B$ \cite{Dadi}, \cite{Firestone}. The charge number
of the final nucleus $\left( ^{7}Be\right) $ is $Z=4$. The full line in Fig.
1. shows $\sigma _{T}=\frac{4\pi }{3}\sigma _{00}$, which is the laser free,
averaged cross section of our model (for $\sigma _{00}$ see $\left( \ref%
{Sigman20}\right) $ with $Q_{0}$). For comparison the $\frac{4\pi }{3}\sigma
_{00}/\chi _{C}^{2}\left( Q_{0}\right) $ is also plotted as a dotted line.
That is the result of the gauge independent laser-free calculation in the
plane wave approximation, i.e. without the Coulomb correction. On the base
of Fig. 1. one can conlude that the Coulomb correction becomes more
essential with decreasing $\gamma $ photon energy.

\begin{figure}[tbp]
\begin{center}
\resizebox{7.5cm}{!}{\includegraphics*{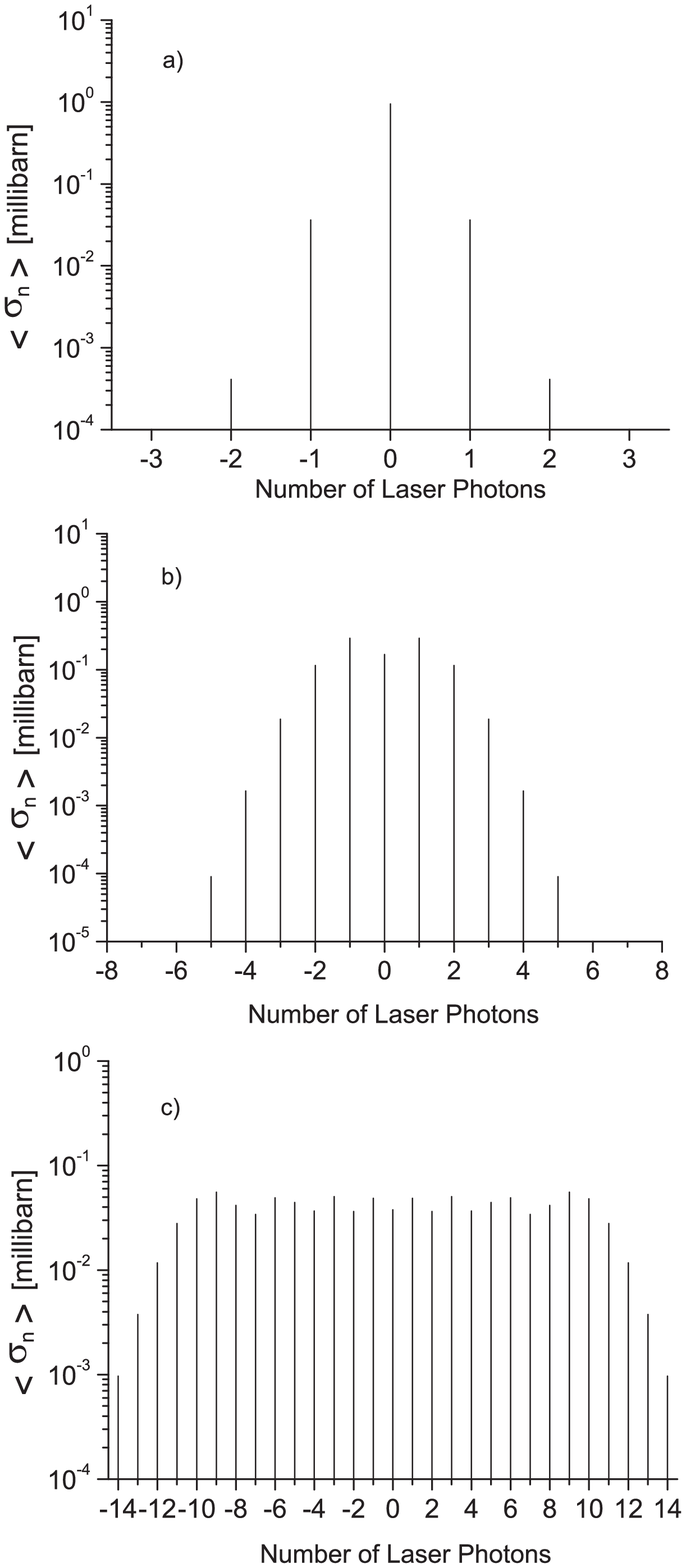}}
\end{center}
\caption{The averaged partial cross sections $\left\langle \protect\sigma %
_{n}\right\rangle$ for some intensity parameters discussed in \protect\cite%
{Dadi}; the intensities are (a) $I=4.0\times 10^{21}(b)1.\times 10^{23}$ and
(c) $2.5\times 10^{24}$ $W/cm^{2}$ with photon energy $\hbar \protect\omega %
_{0}=2 $ $keV$. The polarization vector of the $\protect\gamma $ photon of
energy $3$ $MeV$ is random.}
\label{figure2}
\end{figure}

Next the averaged partial cross sections $\left\langle \sigma
_{n}\right\rangle $ (applying the $\sigma _{n0}=\frac{3}{4\pi }\sigma _{T}$
substitution in $\left( \ref{Signab4}\right) $) are investigated numerically
with $\hbar \omega _{0}=2$ $keV$, $\hbar \omega _{\gamma }=3$ $MeV$ and the
intensities discussed in \cite{Dadi}, i.e. at $I=4.0\times 10^{21},1.\times
10^{23}$ and $2.5\times 10^{24}$ $W/cm^{2}$ (Fig. 2). Figs. 3 and 4 show the
partial cross sections $\sigma _{n,pol,j}$ in the two cases of polarization
of the $\gamma $ photon discussed. Fig. 3 is devoted to the case of $%
\overrightarrow{\varepsilon }=\overrightarrow{e}_{2}$ (case $pol,1$) and
Fig. 4 shows the partial cross sections in the case of $\overrightarrow{%
\varepsilon }=\overrightarrow{e}_{1}\times \overrightarrow{e}_{2}$ (case $%
pol,2$).

\begin{figure}[ptb]
\begin{center}
\resizebox{7.5cm}{!}{\includegraphics*{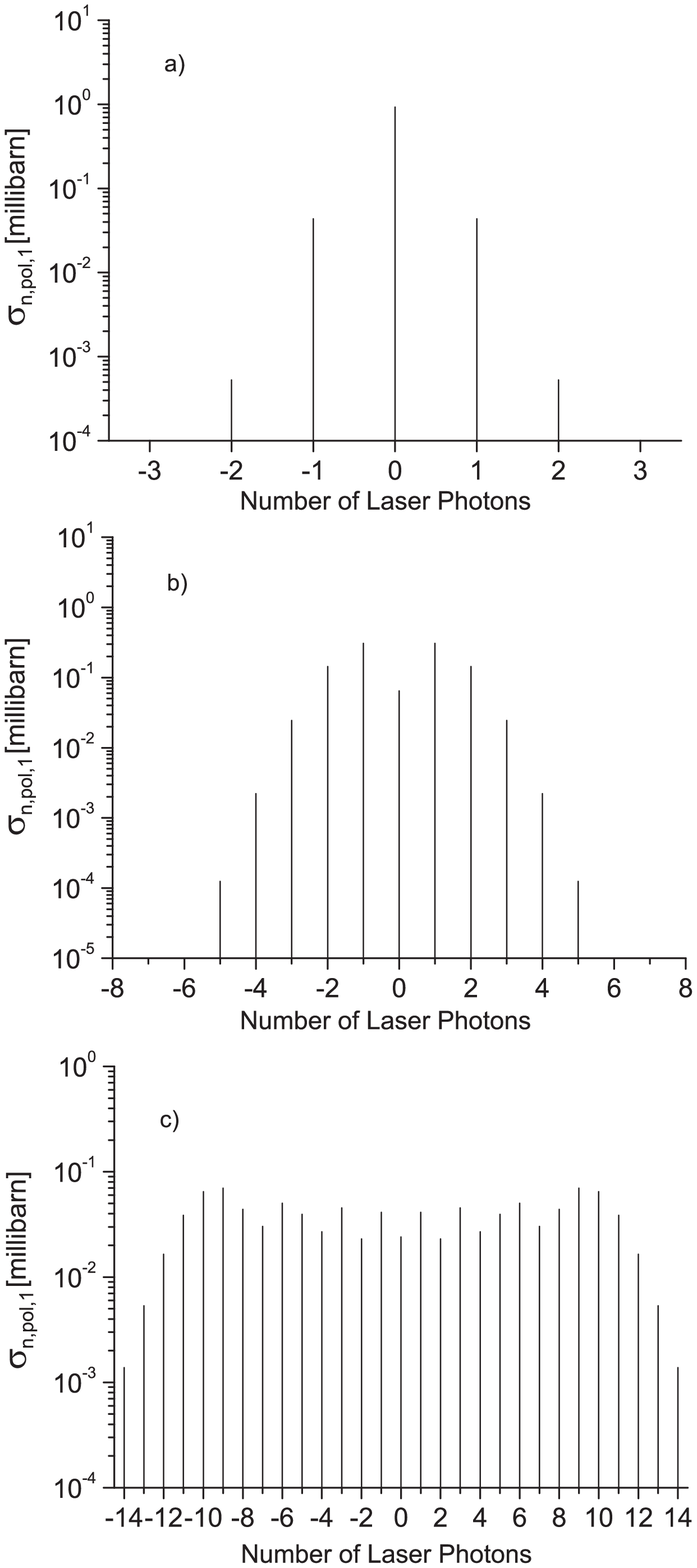}}
\end{center}
\caption{The partial cross sections $\protect\sigma _{n,pol,1}$ in the case
of the state of polarization of the $\protect\gamma $ photon $\protect%
\overrightarrow{\protect\varepsilon }=\protect\overrightarrow{e}_{2}$. The
intensities are (a) $I=4.0\times 10^{21} (b) 1.\times 10^{23}$ and (c) $%
2.5\times 10^{24}$ $W/cm^{2}$ with photon energy $\hbar \protect\omega %
_{0}=2 $ $keV$.}
\label{figure3}
\end{figure}

\begin{figure}[ptb]
\begin{center}
\resizebox{7.5cm}{!}{\includegraphics*{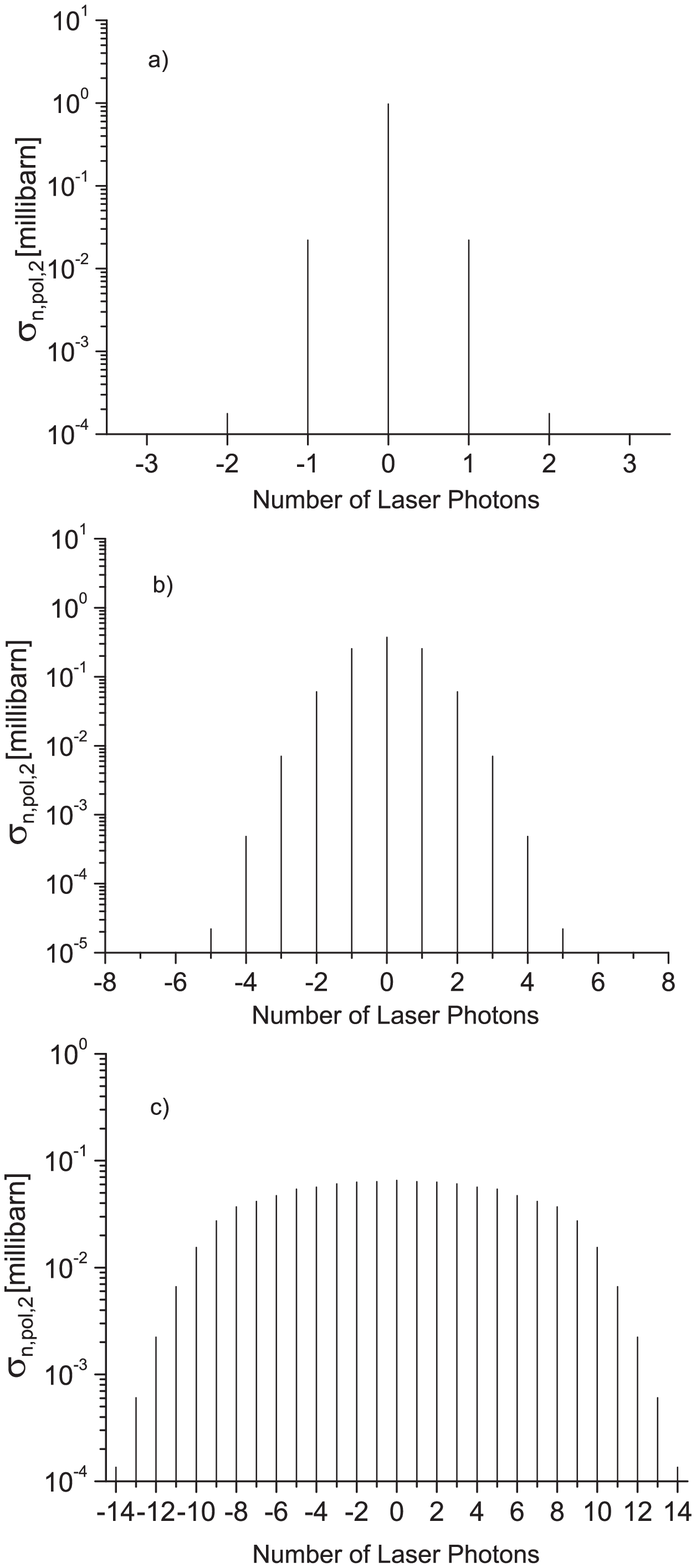}}
\end{center}
\caption{The partial cross sections $\protect\sigma _{n,pol,2}$ in the case
of the state of polarization of the $\protect\gamma $ photon $\protect%
\overrightarrow{\protect\varepsilon }=\protect\overrightarrow{e}_{1}\times 
\protect\overrightarrow{e}_{2}$. The intensities are (a) $I=4.0\times
10^{21} (b) 1.\times 10^{23}$ and (c) $2.5\times 10^{24}$ $W/cm^{2}$ with
photon energy $\hbar \protect\omega _{0}=2 $ $keV$.}
\label{figure4}
\end{figure}

The laser free, averaged total cross section $\sigma _{T}=1.02$ $mb$ at $%
\hbar \omega _{\gamma }=3$ $MeV$. At this $\gamma $ energy and in all cases
we obtained $\left\langle \sigma \right\rangle =\sigma _{pol,1}=$ $\sigma
_{pol,2}=\sigma _{T}=1.02$ $mb$ contrary to $\sigma =\sum_{n}$ $\sigma
_{n}\approx 63.4$ $mb$ obtained by \cite{Dadi}. Also at this $\gamma $
energy the gauge independent laser-free result in the plane wave
approximation gives $\sigma _{T}/\chi _{C}^{2}\left( Q_{0}\right) =4.11$ $mb$%
. Thus one can conclude that the $62.2$ times larger total cross section of 
\cite{Dadi} compared to our result may be originated partly from the gauge
invariant calculation (a factor of $15.4$) and partly from ignoring the
Coulomb repulsion between the final particles by \cite{Dadi} (a factor of $%
4.03$). In the $E_{p}\rightarrow 0$ $\left( \hbar \omega _{\gamma
}\rightarrow 0.137MeV\right) $ limit this type of difference is more
enhanced.

\section{Summary}

The problem of laser-assisted nuclear photo-effect was discussed in a gauge
invariant manner and taking into account the effect of the Coulomb repulsion
between the ejected proton and the remainder nucleus. The investigation is
mainly concerned with those $\gamma $ photon energies that are far from the
threshold. In our model the calculation of the total cross section leads to $%
0.0161$ times smaller result than that of \cite{Dadi} at the $\gamma $\
photon energy $\hbar \omega _{\gamma }=3$ $MeV$ and in all cases of state of
polarization of the $\gamma $ photon discussed. It was found that the
hindering effect of the Coulomb repulsion in the final state, that is
manifested in the appearance of the Coulomb factor in the cross section is
huge for small kinetic energies of the outgoing proton and it must also be
taken into account in the $MeV$ energy range. At $\hbar \omega _{\gamma }=3$ 
$MeV$ it causes about a factor of $1/4$ decrease of the total cross section,
which is incorporated in the factor $0.0161$, compared to the result of the
plane wave approximation. The further $0.065$ times decrease of the total
cross section may be originated from its gauge invariant calculation. The
numerical investigation, similar to \cite{Dadi}, was made at $\hbar \omega
_{\gamma }=3$ $MeV$, and it shows, that the main effect of the presence of
the laser field is that the total cross section is distributed between the
partial cross sections of the channels absorbing or emitting different
numbers of laser photons.

%\section*{References}

\bigskip

\bigskip

\bigskip

\bigskip

\bigskip

\end{document}